\documentclass{sig-alternate}

\usepackage{amsmath,amstext,amssymb,epsf}
\usepackage{epsfig}
\usepackage{verbatim}
\usepackage{pslatex}

\begin{document}
\conferenceinfo{CF'05,} {May 4--6, 2005, Ischia, Italy.}

\CopyrightYear{2005}

\crdata{1-59593-018-3/05/0005}

\bibliographystyle{plain}

\newcommand{\Gn}{{\cal G}_n}
\newcommand{\chooose}[2]{\left( \begin{array}{c} #1 \\ #2 \end{array} \right)}
\newcommand{\Aut}{{\it Aut}}
\newcommand{\Prob}{{\it Prob}}
\newcommand{\Exp}{{\it Exp}}
\newcommand{\Avg}{{\it Avg}}
\newcommand{\In}{\{0,1,\ldots,n-1\}}
\newtheorem{theorem}{\sc Theorem}
\newtheorem{lemma}{\sc Lemma}
\newtheorem{coro}{\sc Corollary}
\newtheorem{nota}{\sc Notation}
\newtheorem{defin}{\sc Definition}
\newtheorem{rem}{\sc Remark}
\newtheorem{cla}{\sc Claim}
\newtheorem{ex}{\sc Example}
\newenvironment{example}{\begin{ex}}{\hspace*{\fill}$\Diamond$\end{ex}}
\newenvironment{claim}{\begin{cla}}{\end{cla}}
\newenvironment{corollary}{\begin{coro}}{\end{coro}}
\newenvironment{definition}{\begin{defin}}{\end{defin}}
\newenvironment{remark}{\begin{rem}}{\end{rem}}
\newenvironment{notation}{\begin{nota}}{\end{nota}}
 
\newcommand{\half}{\frac{1}{2}}
 
\title{Time, Space, and Energy in Reversible Computing}
\numberofauthors{1}
\author{
Paul Vit\'{a}nyi
\titlenote{Part of this work was done while the author was on sabbatical leave
at National ICT of Australia, Sydney Laboratory at UNSW.
Supported in part
by the EU Project RESQ IST-2001-37559,
the ESF QiT Programmme,
the EU NoE PASCAL, and the Netherlands BSIK/BRICKS project.
Address: CWI, Kruislaan 413, 1098 SJ Amsterdam, The Netherlands.
Email: Paul.Vitanyi@cwi.nl}\\
\affaddr{CWI}\\
\affaddr{University of Amsterdam}\\
\affaddr{National ICT of Australia}\\
}

\maketitle
 
\begin{abstract}
We survey results of a quarter century of work on
computation by reversible general-purpose computers (in this
setting Turing machines), and general reversible simulation
of irreversible computations, with respect to
energy-, time- and space requirements.
\end{abstract}

{\bf Categories and Subject Descriptors}: F.2 [Algorithms], F.1.3 [Performance]

{\bf General Terms}: Algorithms, Performance


\section{Introduction}
Computer power
has roughly doubled every 18 months for the last half-century
(Moore's law).
This increase in power is due primarily to the continuing
miniaturization of the elements of which computers are made,
resulting in more and more elementary gates per unit area
 with higher and higher
clock frequency, accompanied by less and less
energy dissipation per elementary computing event. Roughly, a
linear increase in clock speed is accompanied by a square increase
in elements per unit area---so if all elements
compute all of the time, then the dissipated energy per time unit rises
cubicly (linear times square)
in absence of energy decrease per elementary event. The continuing
dramatic decrease in dissipated energy per elementary event is
what has made Moore's law possible. But there is a foreseeable
end to this: There is a
minimum quantum of energy dissipation associated with
elementary events. This puts a fundamental limit
on how far we can go with miniaturization, or does it?

R. Landauer~\cite{La61}
has demonstrated
that it is only the `logically
irreversible' operations in a physical computer
that necessarily dissipate energy by generating a
corresponding amount of entropy for every bit of information
that gets irreversibly erased; the logically reversible operations
can in principle be performed dissipation-free.
One should sharply distinguish between the issue
of logical reversibility and the issue of energy dissipation
freeness.
If a computer operates in a logically
reversible manner, then it still may
dissipate heat.
For such a computer we know that the laws
of physics do not preclude that one can invent a technology
in which to implement a logically similar computer to operate
physically in a dissipationless manner. Computers built
from reversible circuits, or the reversible
Turing machine, \cite{Be73,Be82,FT82}, implemented
with current technology will presumably dissipate energy
but may conceivably be implemented by future technology
in an adiabatic fashion. But for logically irreversible computers
adiabatic implementation is widely considered impossible.

Thought experiments
can exhibit a computer that is both logically and physically
perfectly reversible
and hence perfectly dissipationless. An example is the billiard ball
computer, \cite{FT82}, and similarly the possibility of a
coherent quantum computer, \cite{Fe82,NC00}.
Our purpose is to determine
the theoretical ultimate limits to which the irreversible actions
in an otherwise reversible computation can be reduced.

Currently, computations are commonly irreversible, even though the
physical devices that execute them are fundamentally reversible.
At the basic level, however, matter is governed by classical
mechanics and quantum mechanics,
which are reversible.
This contrast is only possible at the cost of efficiency loss
by generating thermal entropy into the environment.
With computational device technology rapidly approaching
the elementary particle level it has been
argued many times that this effect gains in significance
to the extent that
efficient operation (or operation at all)
of future computers requires them to
be reversible,
\cite{La61,Be73,Be82,FT82,Ke88,LV96,FKM97}.
The mismatch of computing organization and reality
will express itself in friction: computers will 
dissipate a lot of heat unless their mode of operation becomes
reversible, possibly quantum mechanical. 
Since 1940 the
dissipated energy per bit operation in a computing
device has---with remarkable regularity---decreased
at the inverse rate of Moore's law \cite{Ke88} (making Moore's law possible).
Extrapolation of current trends
shows that the energy dissipation per
binary logic operation needs to be reduced below $kT$
(thermal noise)
within 20 years. Here $k$ is Boltzmann's constant and $T$
the absolute temperature in degrees Kelvin,
so that $kT \approx 3 \times 10^{-21}$
Joule at room temperature. Even at $kT$ level,
a future device containing 1 trillion ($10^{12}$) gates
operating at 1 terahertz ($10^{12}$) switching all gates all of the
time dissipates about 3000 watts.
Consequently, in contemporary
 computer and chip architecture design the issue of power consumption
has moved from a background worry to a major problem.
Theoretical advances in reversible
computing are scarce and far between; many serious ones are listed 
in the references. For this review we have drawn primarily
on the results and summarized material in \cite{LV96,LTV98,BTV01}.
It is a tell-tale sign of the difficulty of this area,
that no further advances have been made in this important topic
since that time.

\section{Reversible Turing Machines}
There is a decisive difference between reversible circuits and
reversible special purpose computers \cite{FT82} on the one hand, and
reversible universal computers on the other hand \cite{Be73,Be89}.
While one can design a special-purpose reversible version for every particular
irreversible circuit using reversible universal gates, such a method
does not yield an irreversible-to-reversible compiler
that can execute any irreversible
program on a fixed universal reversible computer architecture as
we are interested in here.

In the standard model of a Turing machine
the elementary operations are rules
in quadruple format $(p,s,a,q)$ meaning that
if the finite control is in state $p$ and the machine
scans tape symbol $s$, then the machine performs action $a$
and subsequently the finite control enters state $q$.
Such an action $a$ consists of either printing a symbol $s'$
in the tape square scanned, or moving the scanning head
one tape square left or
right.

Quadruples are said
to {\em overlap in domain} if they cause the machine in the same state
and scanning the same symbol to perform different actions.
A {\em deterministic Turing machine} is defined as a Turing machine
with quadruples no two of which overlap in domain.

Now consider the special format (deterministic) Turing
machines using quadruples of two
types: {\em read/write} quadruples and {\em move} quadruples.
A read/write quadruple $(p,a,b,q)$ causes the machine in state
$p$ scanning tape symbol $a$ to write symbol $b$ and enter state $q$.
A move quadruple $(p,\ast,\sigma ,q)$ causes the machine
in state $p$ to move its tape head by $\sigma \in \{-1,+1\}$
squares and enter state $q$,
oblivious to the particular symbol in the currently scanned tape square.
(Here `$-1$' means `one square left', 
and `$+1$' means `one square right'.) Quadruples are said
to {\em overlap in range} if they cause the machine to enter
the same state and either both write the same symbol or
(at least) one of them moves the head. Said differently,
quadruples that enter the same state overlap in range
unless they write different symbols.
A {\em reversible Turing machine} is a deterministic Turing machine
with quadruples no two of which overlap in range.
A $k$-tape reversible Turing machine uses $(2k+2)$ tuples
which, for every tape separately, select a read/write or move on that
tape. Moreover, any two tuples can be restricted to some single tape
where they don't overlap in range.

  Simulation of irreversible Turing machines by reversible ones goes back
to Lecerf~\cite{Le63} in 1963, a paper that was little noticed and
only recently rediscovered,  and independently 
Bennett~\cite{Be73} in 1973, which paper
commonly gets all the credit.
To show that every partial recursive function can be computed
by a reversible Turing machine one can proceed as follows \cite{Be73}.
Take the standard irreversible Turing machine computing that function.
We modify it by
adding an auxiliary storage tape called the `history tape'.
The quadruple rules are extended to 6-tuples to additionally
manipulate the history tape.
To be able to reversibly undo (retrace)
the computation deterministically, the new 6-tuple
rules have the effect that the machine keeps a record
on the auxiliary history tape consisting of
the sequence of quadruples executed on the original tape.
Reversibly undoing a computation
entails also erasing the record
of its execution from the history tape.
This notion of reversible computation means
that only $1:1$ recursive functions can be computed.
To reversibly simulate an
irreversible computation from $x$ to $f(x)$
one reversibly  computes from input $x$
to output $\langle x, f(x) \rangle$.

Reversible Turing machines or other reversible computers
will require special reversible programs. One feature of such
programs is that they should be executable when read from bottom
to top as well as when read from top to bottom. Examples are
the programs in the later sections.
In general, writing reversible programs will be difficult.
However, given a general reversible simulation of irreversible computation,
one can simply write an oldfashioned
irreversible program in an irreversible programming language,
and subsequently simulate it reversibly. This leads to the following:
\begin{definition}
An {\em irreversible-to-reversible compiler}
receives an irreversible program as input and
 compiles
it to a reversible program. 
\end{definition}

\section{Adiabatic Computation}
\label{sect.thermo}
All computations can be performed logically reversibly, \cite{Be73},
at the cost of eventually filling up the memory with
unwanted garbage information. This means that
reversible computers with bounded memories
require in the long run irreversible bit operations,
for example, to erase records irreversibly to
create free memory space.
The minimal possible number of irreversibly 
erased bits to do so is believed to determine
the ultimate limit of heat dissipation 
of the computation by Landauer's principle, \cite{La61,Be73,Be82}.
In reference \cite{BGLVZ98} we and others developed a mathematical
theory for the unavoidable number of irreversible bit
operations in an otherwise reversible
computation. 

Methods to implement (almost) 
reversible and dissipationless computation using
conventional technologies
are often designated by the catch phrase `adiabatic switching'.
Many currently proposed physical schemes
implementing adiabatic computation
reduce irreversibility by using longer switching times.
This is done typically by switching over equal voltage gates
after voltage has been equalized slowly. This type of switching
does not dissipate energy, the only energy dissipation
is incurred by pulling voltage up and down: the slower
it goes the less energy is dissipated.
If the computation
goes infinitely slow, zero energy is dissipated. Clearly, this
counteracts the purpose of low energy dissipation
which is faster computation.

In \cite{LV96} it is demonstrated that even if adiabatic computation technology advances
to switching with no time loss,
a similar phenomenon arises when we try to
approach the ultimate limits of minimal irreversibility
of an otherwise reversible computation, and hence minimal
energy dissipation. This time the effect is due to the
logical method of reducing the number of irreversible
bit erasures in the computation
irrespective of
individual switching times.
By computing longer and longer (in
the sense of using more computation steps),
the amount of dissipated energy gets closer to
ultimate limits. Moreover, one can trade-off time (number of steps)
for energy: there is a new
time-irreversibility (time-energy) trade-off hierarchy.
The bounds we derive are also relevant for quantum computations
which are reversible except for the irreversible
observation steps, \cite{Sh94,NC00}.

\subsection{Background}
Around 1940 a computing device dissipated
about $10^{-2}$ Joule per bit operation at room temperature.
Since that time the
dissipated energy per bit operation has roughly decreased
by one order of magnitude (tenfold) every five years.
Currently, a bit operation dissipates\footnote{After
R.W. Keyes, IBM Research.}
about $10^{-17}$ Joule.

Considerations of thermodynamics of
computing started 
in the early fifties. J. von Neumann reputedly thought
that a computer operating at temperature $T$ must
dissipate at least $k T \ln 2$ Joule per elementary bit
operation (about $3 \times 10^{-21}$ J
at room temperature).

Around 1960, R. Landauer~\cite{La61}
more thoroughly analyzed this question
and concluded that it is only `logically 
irreversible' operations
that dissipate energy. An operation is {\it logically reversible}
if its inputs can always be deduced from the outputs.
Erasure of information in a way such that it cannot be retrieved
is not reversible. Erasing each bit costs $k T \ln 2$ energy,
when computer operates at temperature $T$.

\subsection{Irreversibility Cost of Computation}
The ultimate limits of energy dissipation
by computation will be expressed in number of
irreversibly erased
bits.
Hence we consider compactification of records.
In analogy of garbage collection by
a garbage truck, the cost is less if we compact the garbage
before we throw it away.
The ultimate compactification which can be effectively
exploited is expressed in terms of Kolmogorov complexity.

Technically, the {\em  Kolmogorov complexity} of $x$ given $y$ is the length
of the shortest binary program, for the reference universal
prefix Turing machine, that on input $y$ outputs $x$;
it is denoted as $C(x|y)$. For precise definitions, theory and applications,
see \cite{LiVi97}. The Kolmogorov complexity of $x$ is the length
of the shortest binary program with no input that outputs $x$;
it is denoted as $C(x)=C(x|\lambda)$
where $\lambda$ denotes the empty input.
Essentially, the Kolmogorov complexity of a file
is the length of the ultimate compressed version of the file.

Let ${\bf R} = R_1 , R_2 , \ldots$ be a standard
enumeration of reversible Turing machines, \cite{Be73}.
We define $E(\cdot ,\cdot )$ as in \cite{BGLVZ98} (where it is denoted
as $E_3 (\cdot , \cdot)$).

\begin{definition}
The {\em irreversibility cost} 
$E_R (x,y)$ of computing $y$ from $x$
by a reversible Turing machine $R$ is
is 
\[ E_R (x,y) = \min \{|p|+ |q|: R (\langle x,p\rangle)=\langle y,q\rangle \}. \]
We denote the class of all such cost functions
by ${\cal E}$.
\end{definition}

We call an element $E_Q$ of ${\cal E}$ a 
{\em universal irreversibility cost function},
if $Q \in {\bf R}$, and for all $R$ in {\bf R}
\[ E_{Q} (x,y) \leq E_{R} (x,y) + c_{R} ,\]
for all $x$ and $y$, where $c_R$ is a constant which
depends on $R$ but not on $x$ or $y$.
Standard arguments from the theory of Turing machines
show the following. 

\begin{lemma}
There is a universal irreversibility cost function
in ${\cal E}$. Denote it by $E_{UR}$.
\end{lemma}
\begin{proof}
In \cite{Be73} a universal reversible Turing machine $UR$
is constructed which satisfies the optimality requirement.
\end{proof}

Two such universal (or optimal) machines $UR$ and $UR'$ will
assign the same irreversibility cost to a computation
apart from an additive constant term $c$ which is {\em independent}
of $x$ and $y$
(but does depend on $UR$ and $UR'$).
We select a reference universal function $UR$
and define the
{\em irreversibility cost} $E(x,y)$ of
computing $y$ from $x$ as 
\[E(x,y) \equiv  E_{UR} (x,y) . \]
Because of the expression for $E(x,y)$ in Theorem~\ref{thermo.distance}
below it is called the {\em sum cost} measure in \cite{BGLVZ98}.

In physical terms
this cost is in units of $kT \ln 2$, where $k$ is Boltzmann's constant,
$T$ is the absolute temperature in degrees Kelvin,
and $\ln$ is the natural logarithm.

Because the computation is reversible, this definition
is {\em symmetric}: we have $E(x,y)=E(y,x)$.

In our definitions we have pushed all bits to be
irreversibly provided to the start of the computation
and all bits to be erased to the end of the computation.
It is easy to see that this is no restriction. If we have
a computation where irreversible acts happen throughout
the computation, then we can always mark the bits to
be erased, waiting with actual erasure until the end
of the computation.
Similarly, the bits to be provided can be provided
(marked) at the start of the computation while the actual
reading of them (simultaneously unmarking them)
takes place throughout the computation).

Now let us consider a general computation which
outputs string $y$ from input string $x$.
We want to know the
minimum irreversibility cost for such computation.
This leads to the following
theorem, for two different proofs see
\cite{BGLVZ98,LV96}.

\begin{theorem}[Fundamental Theorem]\label{thermo.distance}
Up to an additive logarithmic term
\[E(x,y) = C(x|y)+C(y|x) .\]
\end{theorem}

Erasing a record $x$ is actually a computation from
$x$ to the empty string $\epsilon$. Hence its irreversibility
cost
 is $E(x, \epsilon )$.

\begin{corollary}\label{energy.coro}
Up to a logarithmic additive term,
the irreversibility
cost of erasure is
$E(x, \epsilon )= C(x)$.
\end{corollary}

\subsection{Trading Time for Energy}
Because now the time bounds are
important we consider the universal Turing machine $U$
to be the machine with two work tapes which 
can simulate $t$ steps of a multitape Turing machine $T$
in $O(t \log t)$ steps (the Hennie-Stearns simulation).
If some multitape Turing machine $T$
computes $x$ in time $t$ from a program $p$,
then $U$ computes $x$ in time $O(t \log t)$ from
$p$ plus a description of $T$. 
\begin{definition}
Let $C^t(x|y)$ be the {\em minimal length} of binary program
(not necessarily reversibly) for the 
two work tape universal Turing machine $U$
computing $x$ given $y$ (for free) {\em in time} $t$. Formally,
\[ C^t (x|y) = \min_{p \in {\cal N}}
 \{|p|: U(\langle  p,y \rangle )= x
\mbox{ in $\leq t(|x|)$ steps} \}. \]
$C^t(x|y)$ is called the $t$-{\em time-limited conditional Kolmogorov
complexity} of $x$ given $y$. The unconditional
version is defined as $C^t(x):=C^t(x, \epsilon)$.
A program $p$ such that $U (p)=x$
in $\leq t(|x|)$ steps and $|p|=C^t(x)$ is denoted as $x^*_t$.
\end{definition}

Note that with $C_T^t (x|y)$ the
conditional $t$-time-limited Kolmogorov complexity
with respect to Turing machine $T$, for all $x,y$,
$C^{t'} (x|y) \leq C_T^t (x|y) + c_T$, where $t'=O(t \log t)$ and
$c_T$ is a constant depending on $T$ but not on $x$
and $y$.

This $C^t(\cdot)$ is the standard definition of time-limited
Kolmogorov complexity, \cite{LiVi97}.
However, in the remainder of the paper
we always need to use reversible computations. Fortunately,
in \cite{Be89} it is shown that
for any $\epsilon >0$, ordinary multitape Turing machines
using $T$ time and $S$ space can be simulated by reversible
ones using time $O(T)$ and space $O(ST^{\epsilon})$.

To do effective erasure of compacted information,
we must at the start of the computation provide
a time bound $t$. Typically, $t$ is a recursive function
and the complexity of its description is small, say $O(1)$.
However, in Theorem~\ref{time.1} we allow for very large
running times in order to obtain smaller $C^t(\cdot)$
values.
\begin{theorem}[Effective Erasure]\label{time.1}
If $t(|x|) \geq |x|$ is a
time bound which is provided at the start of the
computation,
then erasing an $n$ bit record $x$ by an otherwise
reversible computation can be done
in time (number of steps) $O(2^{|x|}t(|x|))$
at irreversibility cost (hence also thermodynamic cost)
$C^t (x)+ 2C^t (t|x)+4 \log C^t (t|x)$ bits. (Typically we
consider $t$ as some standard explicit time bound
and the last two terms adding up to $O(1)$.)
\end{theorem}
\begin{proof}
Initially we have in memory
input $x$ and a program $p$ of length $C^t(t,x)$
to compute
reversibly $t$ from $x$. To separate
binary $x$ and binary $p$ we need to encode a delimiter
in at most $2 \log C^t (t|x)$ bits.
\begin{enumerate}
\item
Use $x$ and $p$ to reversibly compute $t$.
Copy $t$ and reverse the computation.
Now we have $x$, $p$ and $t$.
\item
Use $t$ to reversibly dovetail the running
of all programs of length less
than $x$ to find the shortest one halting
in time $t$ with output $x$.
This is $x^*_t$. The computation has produced
garbage bits $g(x,x^*_t)$.
Copy $x^*_t$, and reverse the computation to obtain $x$
erasing all garbage bits $g(x,x^*_t)$.
Now we have $x,p,x^*_t,t$ in memory.
\item
Reversibly compute $t$ from $x$ by $p$, cancel one copy of $t$,
and reverse the computation. Now we have $x,p,x^*_t$
in memory. 
\item
Reversibly cancel $x$ using $x^*_t$ by the standard method,
and then erase $x^*_t$ and $p$ irreversibly.  
\end{enumerate}
\end{proof}

\begin{corollary}\label{cor.Kt}
$E(x, \epsilon) \geq \lim_{t
\rightarrow \infty} C^{t} (x) = C(x)$,
and by Theorem~\ref{thermo.distance}
up to an additional logarithmic term, 
$ E(x, \epsilon)
= C(x)$.
\end{corollary}

Essentially, by spending more time we can
reduce the thermodynamic cost of erasure of $x^*_t$ to its
absolute minimum. In the limit we spend
the optimal value $C(x)$ by erasing $x^*$,
since $\lim_{t \rightarrow \infty} x^*_{t} = x^*$.
This suggests the existence of 
a trade-off hierarchy between time and
energy. The longer  
one reversibly computes to perform final irreversible
erasures, the less bits are erased and energy is dissipated.
This intuitive assertion
will be formally stated and rigourously proved below.

\begin{definition}
Let $UR$ be the reversible version of the two worktape
universal Turing machine, simulating the latter in
linear time by Bennett's result mentioned above.
Let $E^t(x,y)$ be the {\em minimum irreversibility cost} of
an otherwise reversible computation from 
$x$ to $y$ {\em in time} $t$. Formally,
\[ E^t (x,y) = \min_{p,q \in {\cal N}}
 \{|p|+|q|: UR(\langle x,p\rangle)=\langle y,q\rangle
\mbox{ in $\leq t(|x|)$ steps} \}. \]
\end{definition}

Because of the similarity with Corollary~\ref{cor.Kt} ($E(x, \epsilon)$
is about $C(x)$) one is erroneously led to believe that 
$E^t(x, \epsilon) = C^t(x)$ up to a log additive term.
However, the time-bounds introduce many differences.
To reversibly compute $x^*_t$ we may require (because of the
halting problem) at least $O(2^{|x|} t(|x|))$ steps after having
decoded $t$, as indeed is the case in the proof of Theorem~\ref{time.1}.
In contrast, $E^t(x, \epsilon)$ is about
the number of bits erased in an otherwise
reversible computation which uses
at most $t$ steps. Therefore, as far as we know possibly
$C^t (x) \geq E^{t'} (x, \epsilon)$
implies $t' = \Omega ( 2^{|x|}t(|x|))$.
More concretely, it is easy to see that for each
$x$ and $t(|x|) \geq |x|$,
\begin{equation}\label{eq.CE}
E^t(x,\epsilon ) \geq C^t(x) \geq  E^{t'} (x, \epsilon)/2 ,
\end{equation}
with $t'(|x|) = O(t(|x|)$.
Theorem~\ref{time.1} can be restated in terms of $E^t(\cdot)$
as
\[E^{t'} (x, \epsilon) \leq C^t (x)+ 2C^t (t|x)+4 \log C^t (t|x),\]
with $t'(|x|)=O(2^{|x|}t(|x|))$.
Comparing this to the righthand inequality of Equation~\ref{eq.CE}
we have improved the upper bound on erasure cost at the expense of
increasing erasure time. However, these bounds only suggest
but do not actually prove that we
can exchange irreversibility for time.
The following result definitely establishes the existence
of a trade-off, \cite{LV96}.

\begin{theorem}[A Trade-Off Hierarchy]
\label{theorem.hierarchy}
For each large enough $n$
there is a string $x$ of length $n$ and a sequence of $m=\frac{1}{2}
\sqrt n$
time functions $t_1(n) < t_2(n) <  \ldots < t_m(n)$,
such that
\[
E^{t_1}(x,\epsilon ) > E^{t_2}(x,\epsilon ) > \ldots > E^{t_m}(x, \epsilon ).
\]
%Setting $y:=\epsilon$, we have
%a time-irreversibility (=energy) trade-off hierarchy for erasure.
\end{theorem}

In the cost measures like $E^t(\cdot, \cdot)$ we have counted
both the irreversibly provided and the irreversibly erased
bits. But Landauer's principle only charges energy 
dissipation costs for irreversibly erased bits.
It is conceivable that the above results would not hold
if one considers as the cost of erasure of a record only
the irreversibly erased bits. 
However, we have show that Theorem~\ref{theorem.hierarchy}
also holds for Landauer's dissipation measure, \cite{LV96},
in exactly the same form and by almost the same proof.

\section{Outline Simulation Results}
Currently, almost no algorithms and
other programs are designed according to reversible principles
(and in fact, most tasks like computing Boolean functions are inherently
irreversible). To write reversible programs by hand is
unnatural and difficult. The natural way is to
compile irreversible
programs to reversible ones. This raises the question about
efficiency of general reversible simulation of irreversible
computation. Let us briefly summarize the research reviewed
in the next sections. 

Suppose the irreversible computation to be
simulated uses $T$ time and $S$ space.
 A first efficient method was proposed by Bennett \cite{Be89},
but it is space hungry and uses
\footnote{By
judicious choosing of simulation parameters this method can be tweaked to
run in $ST^{1+\epsilon}$ time for every $\epsilon >0$
at the cost of introducing a multiplicative constant depending
on $1/\epsilon$. The complexity analysis of \cite{Be89} 
was completed in \cite{LeSh90}.
} 
 time $ST^{\log 3}$ and
space
$S \log T$. If $T$ is maximal, that is, exponential in $S$,
then the space use is $S^2$. This method can be modeled by a reversible
pebble game. Reference \cite{LTV98} demonstrated
that Bennett's method is optimal
for reversible pebble games and that simulation space can be traded
off against limited erasing.
In \cite{LMT97} it was shown that using a method by Sipser \cite{Si}
one can reversibly simulate using only $O(S)$ extra space but at the cost
of using exponential time. In \cite{FA98} a
relativized separation of reversible and 
irreversible space-time complexity classes is given.

These previous results seem to suggest that a reversible simulation is stuck
with either quadratic space use or exponential time use. 
This impression turns out to be false \cite{BTV01}: 

There is a
tradeoff between time and space which has the exponential time
simulation and the quadratic space simulation as extremes and
for the first time gives a range of simulations using simultaneously
subexponential 
($2^{f(n)}$ is subexponential if $f(n) = o(n)$) time and subquadratic space.
The idea is to use Bennett's pebbling game where the pebble steps
are intervals of the simulated computation that are bridged by
using the exponential simulation method. (It should be noted
that embedding Bennett's pebbling game in the exponential method
gives no gain, and neither does any other iteration of embeddings
of simulation methods.) Careful analysis
shows that the simulation using $k$ pebbles takes
$T' := S 3^k 2^{O(T/2^{k})}$ time and $ S' = O(kS)$ space, and
in some cases the upper bounds are tight. For $k=0$ we
have the exponential time simulation method and for $k= \log T$
we have Bennett's method. Interesting values arise for say

(a) $k= \log \log T$ which yields $T'= S (\log T)^{\log 3} 2^{O(T/\log T)}$
and $S'= S \log \log T  \leq S \log S$; 

(b) for $k= \sqrt{\log T}$
we have $S' = S \sqrt{\log T} \leq S \sqrt{S}$ and 
$T' = S 3^{\sqrt{\log T}}$ $ 2^{O(T/2^{\sqrt{\log T}})}$.

(c)  
Let $T,S,T',S'$ be as above. 
Eliminating the unknown $k$ shows the tradeoff
between simulation time $T'$ and extra
simulation space $S'$:
$ T' = S3^{\frac{S'}{S}}2^{O(T/2^{{\frac{S'}{S}}})} $.

(d) Let $T,S,T',S'$ be as above and let the irreversible
computation be halting and compute a function
from inputs of $n$ bits to outputs. For general reversible simulation 
by a reversible Turing machine
using a binary tape alphabet and a single tape,
$S' \geq n + \log T + O(1)$ and $T' \geq T$.
This lower bound is optimal in the sense that
it can be achieved by simulations
at the cost of using time exponential in $S$.

{\bf Main open problem:}
The ultimate question is whether one can
do better, and obtain improved upper and lower bounds on
the tradeoff between time and space of reversible simulation,
 and in particular whether one can have almost linear time
and almost linear space simultaneously.  

\section{Time Parsimonious Simulation}

\subsection{Reversible Programming}

Reversible Turing machines or other reversible computers
will require special reversible programs. One feature of such
programs is that they should be executable when read from bottom
to top as well as when read from top to bottom. Examples are
the programs we show in the later sections.
In general, writing reversible programs will be difficult.
However, given a general reversible simulation of irreversible computation,
one can simply write an oldfashioned
irreversible program in an irreversible programming language,
and subsequently simulate it reversibly. This leads to the following:
\begin{definition}
An {\em irreversible-to-reversible compiler}
receives an irreversible program as input and
 reversibly compiles
it to a reversible program. Subsequently, the reversible program can be
executed reversibly.
\end{definition}
Note that there is a decisive difference between reversible circuits and
reversible special purpose computers on the one hand, and
reversible universal computers on the other hand.
While one can design a special-purpose reversible version for every particular
irreversible circuit using reversible universal gates, such a method
does not yield an irreversible-to-reversible compiler
that can execute any irreversible
program on a fixed universal reversible computer architecture as
we are interested in here.

\subsection{Initial Reversible Simulations}
The reversible simulation in \cite{Be73} of $T$ steps of an
irreversible computation from $x$ to $f(x)$
reversibly  computes from input $x$ 
to output $\langle x, f(x) \rangle$
in $T' = O(T)$ time.
However, since this reversible simulation at some time instant
has to record the entire
history of the irreversible computation, its space use increases
linearly with the number of simulated steps $T$. That is,
if the simulated irreversible computation uses $S$ space, then
for some constant $c > 1$ the simulation uses
$T'\approx c+cT$ time and $S'\approx c + c(S+T)$ space.
This can be an unacceptable amount of space for many practically
useful computations.

%The question arises whether one can reduce
%the amount of auxiliary space needed by the simulation by a
%more clever simulation method or
%by allowing limited amounts of irreversibility.
%
In \cite{Be89} another elegant simulation technique is devised
reducing the auxiliary storage space.
This simulation does not save the entire history of the irreversible
computation but it breaks up the simulated computation
into segments of about $S$ steps
 and saves in a hierarchical manner {\em checkpoints}
consisting of complete instantaneous descriptions of the
simulated machine (entire tape contents, tape heads positions,
state of the finite control). After a later checkpoint is
reached and saved, the simulating machine reversibly
undoes its intermediate computation, reversibly erasing
the intermediate history and reversibly canceling the previously
saved checkpoint. Subsequently, the computation is resumed from
the new checkpoint onwards.
It turns out that this can be done 
using limited time $T^{\log 3}$ and space $S \log T$.
Fine-tuning the method goes as follows:
The reversible computation simulates $k^n$ segments of length $m$
of irreversible
computation in $(2k-1)^n$ segments of length $\Theta (m+S)$
of reversible computation using
$n(k-1)+1$ checkpoint registers using $\Theta (m+S)$
space each, for every $k,n,m$.

This way it is established that there are various tradeoffs
possible in time-space in between $T'= \Theta (T)$ and
$S' = \Theta (TS)$ at one extreme ($k=1, m=T, n=1$) and (with the corrections
of \cite{LeSh90})
$T' = \Theta (T^{1+\epsilon}/S^{\epsilon} )$
and $S'= \Theta ( c(\epsilon) S(1+ \log T/S))$
with $c(\epsilon)= \epsilon 2^{1/\epsilon}$
for every $\epsilon > 0$, 
using always the same simulation method but with different
parameters $k,n$ where $\epsilon = \log_k (2k-1)$ and $m = \Theta (S)$.
Typically, for $k=2$ we have $\epsilon = \log 3$.
Since for $T > 2^S$ the machine goes into
a computational loop, we always have $S \leq \log T$.
Therefore, 
every irreversible Turing machine
using space $S$ can be simulated by a reversible machine
using space $S^2$ in polynomial time. 
Let us note that it is possible to improve the situation
by reversibly simulating {\em only the irreversible} steps.
Call a quadruple of a Turing machine {\em irreversible} if
its range overlaps with the range of another quadruple.
A step of the computation is {\em irreversible} if
it uses an irreversible quadruple. Let the number of
irreversible steps in a $T$ step computation be denoted by $I$.
Clearly, $I \leq T$. The simulation results
hold with $T$ in the auxiliary space use replaced by $I$.
In particular, $S'=O(S\log I)$. In many computations,
$I$ may be much smaller than $T$. There arises the problem
of estimating the number of irreversible  steps in a
computation. (More complicatedly, one could extend the notion
of irreversible step to those steps which can be reversed on local information
alone. In some cases this is possible
 even when the used quadruple itself was irreversible.)

We  at some point conjectured that {\em all} reversible simulations
of an irreversible computation can
essentially be represented as the pebble game defined below,
and that consequently the lower bound of Corollary~\ref{lem.pebble}
applies to all reversible simulations of irreversible
computations. This conjecture was refuted in
\cite{LMT97} using a technique due to
\cite{Si} to show that 
there exists a general reversible simulation of an
irreversible computation using only order $S$
space at the cost of using a thoroughly 
unrealistic simulation time exponential in $S$.

In retrospect the conjecture was phrased too general: 
it should be restricted to {\em useful} simulations---using
linear or slightly superlinear time and space {\em simultaneously}.
The real question is whether there is a compiler
that takes as input any irreversible algorithm $A$ using $S$ space and
$T$ time and
produces a reversible algorithm $B$ such that $B(x)=A(x)$ for all
input $x$ and using $T'=O(T)$ time and $S'=O(S)$ space.
In the extreme cases of time and space use
this is possible: If $S=\Theta (T)$ then the simulation in \cite{Be73}
does the trick, and if $T= \Theta (2^S)$ then the simulation
of \cite{LMT97} works. For all other cases the pebble game
analysis below has been used
in \cite{FA98} to show that any such simulation, if it exists, cannot 
relativize to oracles, or work in cases where the space bound is much less
than the input length. 
(This is a standard method
of giving evidence
that the aimed-for 
result---here: simulation doesn't exist---is likely to be true in case the
result itself is too hard to obtain.)

\subsection{Reversible Pebbling}
Let $G$ be a linear list of
nodes $\{1,2, \ldots , T_G \}$.
We define a {\em pebble game} on $G$ as follows. The game
proceeds in a discrete sequence of steps of a single {\em player}.
There
are $n$ pebbles which can be put on nodes of $G$.
At any time the set of pebbles is divided in
pebbles on nodes of $G$ and the remaining pebbles which are called
{\em free} pebbles. At every step either an existing
 free pebble can be put
on a node of $G$ (and is thus removed from the free pebble pool)
 or be removed from a node of $G$ (and is added to the
free pebble pool).
Initially $G$ is unpebbled and there is a pool of free pebbles.
The game is played according to the following rule:

\begin{description}
\item[Reversible Pebble Rule:]
If node $i$ is occupied by a pebble, then one may either
place a free pebble on node $i+1$ (if it was not occupied before), or
remove the pebble from node $i+1$.
\end{description}

We assume an extra initial node $0$ permanently
occupied by an extra, fixed pebble,
so that node $1$ may be (un)pebbled at will.
This pebble game is inspired by the method of simulating irreversible Turing
Machines on reversible ones in a space efficient manner. The placement
of a pebble corresponds to checkpointing the next state of the irreversible
computation, while the removal of a pebble corresponds to reversibly erasing
a checkpoint. Our main interest is in determining the number of pebbles $k$
needed to pebble a given node $i$.

The maximum number $n$ of pebbles
which are simultaneously on $G$
at any one time in the game gives the space complexity
$nS$ of the simulation. If one deletes a pebble not following
the above rules, then this means a block of bits of size $S$ is
erased irreversibly. 
This pebble game is inspired by the method of simulating irreversible Turing
Machines on reversible ones in a space efficient manner. The placement
of a pebble corresponds to checkpointing the current state of the irreversible
computation, while the removal of a pebble corresponds to reversibly erasing
a checkpoint. Our main interest is in determining the number of pebbles $k$
needed to pebble a given node $i$.

The maximum number $n$ of pebbles
which are simultaneously on $G$
at any one time in the game gives the space complexity
$nS$ of the simulation. If one deletes a pebble not following
the above rules, then this means a block of bits of size $S$ is
erased irreversibly. The limitation to
Bennett's simulation is in fact space, rather than time.
When space is limited, we may not have enough place to store garbage,
and these garbage bits will have to be irreversibly erased.
We establish a tight lower bound for {\em any}
strategy for the pebble game in order to obtain
a space-irreversibility tradeoff.

\subsection{Algorithm}

\label{reachable}

We describe the idea of Bennett's simulation \cite{Be89}.
This simulation is optimal \cite{LTV98} among all reversible pebble games,
and we will show the proof below.
The total computation of $T$ steps is broken into $2^k$ segments of
length $m=T2^{-k}$.
Every $m$th point of the computation is a node in the pebbling game;
node $i$ corresponding to $im$ steps of computation.

For each pebble a section of tape is reserved long enough to store the
whole configuration of the simulated machine. By enlarging the tape
alphabet, each pebble will require space only $S+O(1)$.

%% but enlarging alphabet means that simulation uses O(S) bits!

Both the pebbling and unpebbling of a pebble $t$ on some node,
given that the previous node has a pebble $s$ on it, will be achieved
by a single reversible procedure bridge($s,t$). This looks
up the configuration at section $s$, simulates $m$ steps of computation
in a manner described in section~\ref{bridge}, and exclusive-or's the result
into section $t$.
If $t$ was a free pebble, meaning that its tape section is all zeroes,
the result is that pebble $t$ occupies the next node. If $t$ already
pebbled that node then it will be zeroed as a result.

The essence of Bennett's simulation is a recursive subdivision
of a computation path into 2 halves, which are traversed in 3 stages;
the first stage gets the midpoint pebbled, the second gets the
endpoint pebbled, and the 3rd recovers the midpoint pebble.
The following recursive procedure implements this scheme;
Pebble($s,t,n$) uses free pebbles $0,\ldots,n-1$ to compute
the $2^n$th node after the one pebbled by $s$,
and exclusive-or's that node with pebble $t$
(either putting $t$ on the node or taking it off).
Its correctness follows by straightforward induction.
Note that it is its own reverse;
executing it twice will produce no net change.
The pebble parameters $s$ and $t$ are simply numbers
in the range $-1,0,1,\ldots,k$. Pebble -1 is permanently on node 0,
pebble $k$ gets to pebble the final node, and pebble $i$,
for $0\leq i < k$ pebbles nodes that are odd multiples of $2^i$.
The entire simulation is carried out with a call pebble($-1,k,k$).

%Pebbles $0$ and $n$ are special and
%correspond to the initial and final configuration which use the
%input and output tape instead of a section of the work tape.

\begin{tabbing}
pe\=bb\=le($s,t,n$) \\
\{ \\
\> if ($n=0$) \\
\> \> bridge($s,t$); \\
\> fi ($n=0$) \\
\> if ($n>0$) \\
\> let $r=n-1$ \\
\> pebble($s,r,n-1$); \\
\> pebble($r,t,n-1$); \\
\> pebble($s,r,n-1$) \\
\> fi ($n>0$) \\
\}
\end{tabbing}

As noted by Bennett, both branches and merges must be labeled with
mutually exclusive conditions to ensure reversibility.
Recursion can be easily implemented reversibly by introducing an extra
stack tape, which will hold at most $n$ stack frames of size $O(\log n)$ each,
for a total of $O(n \log n)$.

\subsection{Optimality}
This pebbling method is optimal in that no more
than $2^{n+1}-1$ steps can be bridged with $n$ pebbles \cite{LTV98}.
It is easy to see that the method achieves this;
the difficult part is showing that no reversible pebbling method can do better.
It turns out that characterizing the maximum node that can be pebbled with
a given number of pebbles is best done by completely characterizing
what pebble configurations are realizable.
First we need to introduce some helpful notions.

In a given pebble configuration with $f$ free pebbles,
a placed pebble is called {\em available} if there is another pebble at most
$2^f$ positions to its left ($0$ being the leftmost node).
According to the above procedures, an available pebble can be removed
with the use of the free pebbles. For convenience we imagine this as
a single big step in our game.

Call a pebble configuration {\em weakly solvable} if there is a way
of repeatedly removing an available pebble until all are free.
Note that such configurations are necessarily realizable, since the removal
process can be run in reverse to recreate the original configuration.
Call a pebble configuration {\em strongly solvable} if all ways of repeatedly
removing an available pebble lead to all being free. Obviously any
strongly solvable configuration is also weakly solvable.

The starting configuration is obviously both weakly and strongly
solvable. How does the single rule of the game affect solvability?
Clearly, adding a pebble to a weakly solvable configuation yields
another weakly solvable configuation, while removing a pebble from a strongly
solvable configuation yields another strongly solvable configuation.
It is not clear if removing a pebble from a weakly solvable configuation
yields another one. If such is the case then we may conclude that all
realizable configurations are weakly solvable and hence the two classes
coincide. This is exactly what the next theorem shows.
 
\begin{theorem}
Every weakly solvable configuration is strongly solvable.
\end{theorem} 

\begin{proof}
Let $f$ be the number of free pebbles in a weakly solvable
configuration.
%It suffices to show that removal of any available pebble yields another
%weakly solvable configuration.
Number the placed pebbles $f,f+1,\ldots,n-1$ according
to their order of removal. It is given that, for all $i$,
pebble $i$ has a higher-numbered pebble at most $2^i$ positions
to its left (number the fixed pebble at $0$ infinity). We
know that pebble $f$ is available. Suppose
a pebble $g$ with $g>f$ is also available---so there must be a pebble
at most $2^f$ positions to its left.
It suffices to show that if pebble $g$ is removed first, then pebbles
$f,f+1,\ldots,g-1$ are still available when their turn comes.
Suppose pebble $j$ finds pebble $g$ at most $2^j$ places to its left 
(otherwise $j$ will still be available after $g$'s removal for sure).
Then after removal of pebbles $g,f,f+1,\ldots,j-1$, it will still find
a higher-numbered pebble at most $2^j + 2^f + 2^f + 2^{f+1} + \cdots + 2^{j-1}
\leq 2^{j+1}$ places to its left, thus making it available given the extra
now free pebble $g$.
\end{proof}

\begin{corollary}
A configuration with $f$ free pebbles
is realizable if and only if its placed pebbles can be
numbered $f,f+1,\ldots,n-1$ such that
pebble $i$ has a higher-numbered pebble at most $2^i$ positions
to its left.
\end{corollary}

\begin{corollary}\label{lem.bennett}\label{lem.pebble}
The maximum reachable node with $n$ pebbles is $\sum_{i=0}^{n-1} 2^i = 2^n-1$.
\end{corollary}

Moreover, if pebble($s,n$) takes $t(n)$ steps
we find $t(0) = 1$ and
$t(n)=3 t(n-1) + 1 = (3^{n+1}-1)/2$. That is, the number
of steps $T_G'$ of a winning play of a pebble game
of size $T_G=2^n-1$ is $T_G' \approx 1.5 3^n$, that is,
$T_G' \approx T_G^{\log 3}$.

\subsection{Trading Space for Erasures}
The simulation above follows the rules
of the pebble game of length $T_G = 2^n-1$ with $n$ pebbles above.
A winning
strategy for a game of length $T_G$ using $n$ pebbles
corresponds with reversibly simulating $T_G$ segments of $S$
steps of an irreversible computation using $S$
space such that the reversible simulator
uses $T' \approx ST'_G \approx ST_G^{\log 3}$ steps
and total space $S'=nS$. The space $S'$ corresponds
to the maximal number of pebbles on $G$
at any time during the game.  The placement or removal of a
pebble in the game corresponds to the reversible
copying or reversible cancelation of a `checkpoint'
consisting of the entire instantaneous description of size $S$
(work tape contents, location of heads, state of finite
control) of the simulated irreversible machine.
The total time $T_GS$ used by the irreversible computation
is broken up in segments of size $S$ so that the reversible
copying and canceling of a checkpoints takes about the same
number of steps as the computation segments in between
checkpoints.
\footnote{If we are to account for the permanent pebble on node $0$,
we get that the simulation uses $n+1$
pebbles for a pebble game with $n$ pebbles of length $T_G+1$.
The simulation uses $n+1 =S'/S$ pebbles for
a simulated number of $S(T_G+1)$ steps of the irreversible
computation.}

We can now formulate a tradeoff between space used
by a polynomial time reversible computation and irreversible
erasures as proposed in \cite{LTV98}. First we show that allowing a limited
amount of erasure in an otherwise
reversible computation means that 
we can get by with less work space.
Therefore, we define an {\em $m$-erasure} pebble game as
the pebble game above but with the additional rule

\begin{itemize} 
\item
In at most $m$ steps
the player can
remove a pebble from any node $i > 1$ without
node $i-1$ being pebbled at the time.
\end{itemize} 
  
An $m$-erasure pebble game corresponds with an otherwise
reversible computation using $mS$ irreversible bit erasures,
where $S$ is the space used by the irreversible computation
being simulated.

\begin{lemma}\label{lem.erasure}
There is a winning strategy with $n+2$ pebbles
and $m-1$ erasures for pebble games $G$
with $T_G= m2^n$, for all $m \geq 1$. 
\end{lemma}
\begin{proof}
The strategy is to use 2 pebbles as springboards that are alternately
placed $2^n$ in front of each other using the remaining $n$ pebbles
to bridge the distance. The most backward springboard can be erased
from its old position once all $n$ pebbles are cleared from the space
between it and the front springboard.

The simulation time $T'_G$ is 
$T'_G \approx 2m\cdot 3^{n-1} +2
\approx 2m ( T_G/m)^{\log 3} = 2m^{1- \log 3 } T_G^{\log 3}$
for $T_G = m2^{n-1}$.
\end{proof}

\begin{theorem}[Space-Irreversibility]
\label{theo.si}
(i) Pebble games $G$ of size $2^n-1$ can be won using $n$ pebbles
but not using $n-1$ pebbles.

(ii) If $G$ is a pebble game with a winning strategy
using $n$ pebbles without
erasures, then there is also a winning strategy for $G$
using $E$ erasures and $n-\log (E+1)$ pebbles (for $E$ is an odd
integer at least 1).

\end{theorem}
\begin{proof}
(i) By Corollory~\ref{lem.bennett}.

(ii) By (i), $T_G = 2^n-1$ 
is the maximum length of a pebble game $G$
for which there is a winning strategy using $n$
pebbles and no erasures.
By Lemma~\ref{lem.erasure}, we can pebble a game $G$
of length $T_G= m2^{n-\log m}=2^n$ using $n+1-\log m$
pebbles and $2m-1$ erasures.
\end{proof}

We analyze the consequences of Theorem~\ref{theo.si}.
 It is convenient
to consider the special sequence of values
$E :=2^{k+2}-1$ for $k:=0,1, \ldots$.
Let $G$ be Bennett's pebble game of Lemma~\ref{lem.bennett}
of length $T_G=2^{n}-1$. 
It can be won using $n$ pebbles
without erasures, or using
$n-k $ pebbles plus $2^{k+2}-1$ erasures (which gives a gain
over not erasing as in Lemma~\ref{lem.bennett} only for $k \geq 1$), but not
using $n-1$ pebbles.

Therefore, we can exchange space use
for irreversible erasures.
Such a tradeoff can be used to reduce
the space requirements of the reversible simulation.
The correspondence between the
erasure pebble game and the
otherwise reversible computations
using irreversible erasures is
that if the pebble game uses $n-k$ pebbles
and $2^{k+2} -1$ erasures, then the otherwise reversible
computation uses $(n-k)S$ space and erases $(2^{k+2}-1)S$ bits
irreversibly.

Therefore, a reversible simulation according to the pebble game
 of every irreversible
computation of length $T=(2^n-1)S$ can be done using
$nS$ space using $(T/S)^{\log 3 } S$ time,
 but is impossible using $(n-1)S$ space. It can also
be performed using $(n-k)S$ space, $(2^{k+2}-1)S$
irreversible bit erasures and
 $2^{(k+1)(1-\log 3 )+1} (T/S)^{\log 3} S$
time. In the extreme case
we use no space to store the history and erase about $4T$
bits. This corresponds to the fact that an irreversible
computation may overwrite its scanned symbol irreversibly
at each step.

\begin{definition}
\rm
Consider a simulation according to the pebble game
using $S'$ storage space
and $T'$ time
which reversibly computes
$y = \langle x, f(x) \rangle$ from $x$ in order to
simulate
an irreversible computation
using $S$ storage space and $T$ time
which computes $f(x)$ from $x$.
The {\em irreversible simulation
cost} $B^{S'}(x, y)$ of the simulation
is the number of
irreversibly erased bits in the simulation (with the parameters $S,T,T'$
understood).
\end{definition}

If the irreversible 
simulated computation from $x$ to $f(x)$ uses $T$ steps, then for $S' = nS$
and $n =  \log (T/S)$ we have above treated the most space
parsimonious simulation which yields $B^{S'} (x,y) = 0$, with
$y= \langle x,f(x) \rangle$.

\begin{corollary}
Simulating a $T=(2^{n}-1)S$ step
irreversible computation from $x$ to $f(x)$  
using $S$ space
 by
a computation from $x$ to $y = \langle x, f(x) \rangle$, the
irreversible simulation cost satisfies:

(i) $B^{(n- k)S } (x,y) \leq B^{nS}(x, y) + (2^{k+2}-1)S$,
for $n \geq k \geq 1$. 

(ii) $B^{(n-1)S}(x,y) > B^{nS}(x,y)$, for $n \geq 1$.

\end{corollary} 

For the most space parsimonious
simulation with $n=\log (T/S)$ this means that
\[ B^{S(\log (T/S) - k) } (x,y) \leq
B^{S \log (T/S)}(x, y) + (2^{k+2}-1)S.\]

\section{Space Parsimonious Simulation}
\label{bridge}
Lange, McKenzie and Tapp, \cite{LMT97}, devised a reversible simulation,
{\em LMT-simulation} for short,
that doesn't use extra space, at the cost of using exponential time.
Their main idea of reversibly simulating a machine without using more space
is by reversibly cycling through the configuration tree of the machine
(more precisely the connected component containing the input configuration).
This configuration tree is a tree whose nodes
are the machine configurations and where two nodes are connected
by an edge if the machine moves in one step from one configuration
to the other. We consider each edge to consist of two {\em half-edges},
each adjacent to one configuration.

% Under certain easily satisfied conditions of the simulated machine,
% they showed how to traverse its
The configuration tree can be traversed by alternating
two permutations on half-edges: a swapping permutation which
swaps the two half-edges constituting each edge, and a rotation
permutation whose orbits are all the half-edges adjacent to one
configuration. Both permutations can be implemented in a constant number
of steps. For simplicity one assumes the simulated machine
strictly alternates moving and writing transitions.
To prevent the simulation from exceeding the available space $S$,
each pebble section is marked with special left and right
markers $\dagger, \ddagger$, which we assume the simulated machine not
to cross. Since this only prevents crossings in the forward simulation,
we furthermore, with the head on the left (right) marker,
only consider previous moving transitions from the right (left).

\section{Time--Space Tradeoff}
A call pebble($s,t,n$) results in $3^n$ calls to bridge($\cdot$,$\cdot$).
Bennett chose the number of pebbles large enough ($n=\Omega(\log T)$)
so that $m$ becomes small, on the order of the space $S$ used by the simulated
machine. In that case bridge($s,t$) is easily implemented with the help
of an additional {\em history} tape of size $m$ which records the
sequence of transitions. Instead, \cite{BTV01} showed how to
allow an arbitrary choice of $n$
and resort to the space efficient simulation of \cite{LMT97} to
bridge the pebbled checkpoints. 
A similar solution was arrived at later and independently
in the unpublished manuscript \cite{Wi00}.

To adapt the LMT simulation to our needs, we equip our simulating machine
with one extra tape to hold the simulated configuration and
another extra tape counting the difference between forward and backward
steps simulated. $m=2^n$ steps of computation can be bridged with a $\log m$
bits binary counter, incremented with each simulated forward step,
and decremented with each simulated backward step.
%incurring an extra $O(\log m)$ slowdown in simulation speed.
Having obtained the configuration $m$ steps beyond that of pebble $s$,
it is exclusive-or'd into section $t$ and then the LMT simulation is
reversed to end up with a zero counter and a copy of section $s$, which
is blanked by an exclusive-or from the original.

\begin{tabbing}
br\=id\=ge\=($s,t$) \\
\{ \\
\> copy section $s$ onto (blanked) simulation tape \\
\> setup: goto enter; \\
\> loop1: come from endloop1; \\
\> simulate step with swap\&rotate and adjust counter \\
\> if (counter=0) \\
\> \> rotate back; \\
\> \> if (simulation tape = section $s$) \\
\> \> \> enter: come from start; \\
\> \> fi (simulation tape = section $s$) \\
\> fi (counter=0) \\
\> endloop1: if (counter!=$m$) goto loop1; \\
\> exclusive-or simulation tape into section $t$ \\
\> if (counter!=$m$) \\
\> \> loop2: come from endloop2; \\
\> reverse-simulate step with anti-rotate\&swap and adjust counter \\
\> if (counter=0) \\
\> \> rotate back; \\
\> \> if (simulation tape = section $s$) goto exit; \\
\> fi (counter=0) \\
\> endloop2: goto loop2; \\
\> exit: clear simulation tape using section $s$ \\
\}
\end{tabbing}

\subsection{Complexity Analysis}
Let us analyze the time and space used by this simulation.
\begin{theorem}
An irreversible computation using time $T$ 
and space $S$ can be simulated reversibly
in time $T'=  3^k 2^{O(T/2^k)}S$ and space $S' = S(1+O(k))$,
where $k$ is a parameter that can be chosen freely $0 \leq k \leq \log T$
to obtain the required tradeoff between reversible time $T'$
and space $S'$.
\end{theorem}
\begin{proof} 
(Sketch)
%Suppose we simulate $T$ steps of the irreversible computation.
Every invocation of the bridge() procedure
 takes time $O(2^{O(m)}S)$.
That is, every configuration has at most $O(1)$ predecessor
configurations where it can have come from (constant number of states,
constant alphabet size and choice of direction).
Hence there are
$\leq 2^{O(m)}$ configurations to be searched
and about as many potential start configurations leading in $m$ moves
to the goal configuration, and every tape section comparison takes
time $O(S)$.
The pebbling game over $2^k$ nodes takes $3^k$ (un)pebbling steps
each of which is an invocation of bridge().
Filling in $m=T/2^k$ gives the claimed time bound.
Each of the $k+O(1)$ pebbles takes space $O(S)$,
as does the simulation tape and the counter,
giving the claimed total space.
\end{proof}

It is easy to verify that
for some simulations the upper bound
is tight.
The boundary cases, $k=0$ gives the LMT-simulation using exponential
time and no extra space, and $k= \log T$ gives Bennett's simulation
using at most square space and subquadratic time. Taking intermediate
values of $k$ we can choose to reduce time at the cost of
an increase of space use and vice versa.
In particular, special values $k= \log \log T$ and
$k= \sqrt{T}$ give the results using simultaneously subexponential
time and subquadratic space exhibited in the introduction.
Eliminating $k$ we obtain:

\begin{corollary}
Let $T,S,T',S'$ be as above. Then there is a reversible simulation that
has the following tradeoff between simulation time $T'$ and extra
simulation space $S'$:
\[ T' = S3^{\frac{S'}{S}}2^{O(T/2^{{\frac{S'}{S}}})} . \]
\end{corollary}

\subsection{Local Irreversible Actions}
Suppose we have an otherwise reversible computation containing
local irreversible actions. Then we need to reversibly simulate
only the subsequence of irreversible steps, leaving the connecting reversible
computation segments unchanged.  That is, an irreversiblity 
parsimonious
computation is much cheaper to reversibly simulate than an
irreversibility hungry one.

\section{Unknown Computing Time}
In the previous analysis we have tacitly assumed that the
reversible simulator knows in advance the number of steps $T$ taken
by the irreversible computation to be simulated.
In this context one can distinguish on-line computations
and off-line computations to be simulated. On-line computations
are computations which interact with the outside environment
and in principle keep running forever. An example
is the operating system of a computer.
Off-line computations are computations which compute a definite
function from an input (argument) to an output (value). For example,
given as input a positive integer number, compute as output all
its prime factors. For every input such an algorithm will have a definite
running time.

There is a well-known simple device %(used in detail in \cite{Be89})
to remove this dependency for batch computations
 without increasing the simulation time  (and space)
too much \cite{LTV98}. Suppose we want to simulate a computation with
unknown computation time $T$. Then we simulate $t$ steps
of the computation with $t$ running through the sequence
of values $2,2^{2},2^{3}, \ldots$ For every value $t$ takes on
we reversibly simulate the first $t$ steps of
the irreversible computation. If $T>t$ then the computation is not finished
at the end of this simulation. Subsequently we reversibly undo
the computation until the initial state is reached again, set $t:=2t$
and reversibly simulate again. This way we continue until $t\geq T$
at which bound the computation finishes. The total time spent
in this simulation is 
\[ T''  \leq  2 \sum_{i=1}^{\lceil \log T \rceil} 
 S3^{\frac{S'}{S}}2^{O(2^{i - \frac{S'}{S}})}
 \leq 2 T' .\]

\section{Lower Bound}\label{sect.lb}
It is not difficult to show a simple lower bound \cite{BTV01} on the extra
storage space required for general
reversible simulation. We consider only irreversible computations
that are halting computations performing a mapping
from an input to an output. For convenience 
we assume
that the Turing machine has a single
binary work tape 
delemited by
markers $\dagger, \ddagger$ that are placed $S$ positions apart.
Initially the binary input of length $n$
is written left adjusted on the work tape.
At the end of the computation the output
is written left adjusted on the work tape.
The markers are never moved.
Such a machine clearly can perform every computation as long as $S$
is large enough with respect to $n$.
Assume that the reversible simulator is a similar model albeit reversible.
The average number of steps
in the computation is the uniform average over all equally likely
inputs of $n$ bits.
\begin{theorem}\label{theo.lb}
To generally simulate an irreversible halting computation of
a Turing machine as above using storage
space $S$ and $T$ steps on
average, on inputs of length $n$,
by a general reversible computation
using $S'$ storage space 
and $T'$ steps on average, 
the reversible simulator Turing machine
having $q'$ states, requires trivially $T' \geq T$ and
$S'   \geq n + \log T - O(1)$ up to a logarithmic additive term.
\end{theorem}

\begin{proof}
There are $2^n$ possible inputs to the irreversible computation,
the computation on every input using on average $T$ steps.
A general simulation of this machine cannot use the semantics
of the function being simulated but must simulate every
step of the simulated machine. Hence $T' \geq T$.
 The simulator being reversible
requires different configurations for every step of everyone
of the simulated computations
that is, at least $2^nT$ configurations. 
The simulating machine has not more than $ q'2^{S'} S'$ distinct
configurations---$2^{S'}$ distinct values
on the work tape, $q'$ states, 
and $S'$ head positions for the combination
of input tape and work tape. Therefore, 
$ q'2^{S'} S' \geq 2^nT$. That is,
$q'S'2^{S'-n} \geq T$ which shows that $S' - n - \log S' \geq \log T - \log q'$.
\end{proof}

For example, consider irreversible computations that don't use
extra space apart from the space to hold the input, that is, $S=n$. An
example is the computation of $f(x)=0$. 
\begin{itemize}
\item
If $T$ is polynomial in $n$ then $S' = n+ \Omega (\log n)$.
\item
If $T$ is exponential in $n$ then $S' = n+ \Omega (n)$.
\end{itemize}

Thus, in some cases the LMT-algorithm is required
to use extra space if we
deal with halting computations computing a function from
input to output. In the final version of the paper \cite{LMT97} the authors
have added that their simulation uses some extra space for counting
(essentially $O(S)$) in case we require halting computations
from input to output, matching the lower bound above for $S=n$ since their 
simulation uses on average $T'$ steps exponential in $S$. 

{\bf Optimality and Tradeoffs:} 
The lower bound of Theorem~\ref{theo.lb} is optimal in the following sense.
As one extreme, the LMT-algorithm of \cite{LMT97}
discussed above uses $S' = n+ \log T$  space
for simulating irreversible computations of total functions on
inputs of $n$ bits, but
at the cost of using $T' = \Omega ( 2^S)$ simulation time. As the other
extreme, Bennett's simple algorithm in \cite{Be73} uses $T' = O(T)$
reversible simulation time, but at the cost of using $S' = \Omega (T)$
additional storage space. This implies that improvements in determining
the complexity of reversible simulation must consider time-space tradeoffs.

\end{document}